\documentclass[conference]{IEEEtran}
\IEEEoverridecommandlockouts

\makeatletter
\def\ps@headings{%
\def\@oddhead{\mbox{}\scriptsize\rightmark \hfil \thepage}%
\def\@evenhead{\scriptsize\thepage \hfil \leftmark\mbox{}}%
\def\@oddfoot{}%
\def\@evenfoot{}}
\makeatother

\usepackage[left=0.625in,right=0.625in,top=0.8in, bottom=1.04in]{geometry}

\usepackage{algorithm}
\usepackage{algorithmicx}
\usepackage{algpseudocode}
\usepackage{amsmath,amssymb,amsfonts}
\usepackage{amsthm}
\usepackage{graphicx}
\usepackage{textcomp}
\usepackage{mathptmx}
\usepackage{amsmath}
\usepackage{tabularx}
\usepackage{subfigure}
\usepackage{xcolor}
\usepackage{multirow}
\usepackage{url}
\usepackage{hyperref}
\usepackage{booktabs}
\usepackage{caption}
\usepackage{makecell}
\usepackage{tikz}
\usetikzlibrary{quantikz2}

\usepackage{url}
\usepackage{algorithm,algpseudocode}

\def\BibTeX{{\rm B\kern-.05em{\sc i\kern-.025em b}\kern-.08em
    T\kern-.1667em\lower.7ex\hbox{E}\kern-.125emX}}

\begin{document}

\title{\huge{SQUASH: A SWAP-Based Quantum Attack to Sabotage Hybrid Quantum Neural Networks}}

\author{Rahul Kumar$^*$, Wenqi Wei$^*$, Ying Mao$^*$, Junaid Farooq$^\dagger$, Ying Wang$^\ddagger$, and Juntao Chen$^*$
\thanks{$^*$The authors are with the Department of Computer and Information Sciences, Fordham University, New York, NY, 10023 USA. E-mail: \{rkumar42, wwei23, ymao41, jchen504\}@fordham.edu}
\thanks{$^\dagger$The author is with the Department of Electrical
\& Computer Engineering, College of Engineering and Computer Science, University of Michigan-Dearborn, Dearborn, MI 48128 USA. E-mail: mjfarooq@umich.edu}
\thanks{$^\ddagger$The author is with the Department of Systems and Enterprises, Stevens Institute of Technology, Hoboken, NJ 07030 USA. E-mail: ywang6@stevens.edu}
}

\maketitle

\begin{abstract}
We propose a circuit-level attack, SQUASH, a SWAP-Based Quantum Attack to sabotage Hybrid Quantum Neural Networks (HQNNs) for classification tasks. SQUASH is executed by inserting SWAP gate(s) into the variational quantum circuit of the victim HQNN. Unlike conventional noise-based or adversarial input attacks, SQUASH directly manipulates the circuit structure, leading to qubit misalignment and disrupting quantum state evolution. This attack is highly stealthy, as it does not require access to training data or introduce detectable perturbations in input states. Our results demonstrate that SQUASH significantly degrades classification performance, with untargeted SWAP attacks reducing accuracy by up to 74.08\% and targeted SWAP attacks reducing target class accuracy by up to 79.78\%. These findings reveal a critical vulnerability in HQNN implementations, underscoring the need for more resilient architectures against circuit-level adversarial interventions.
\end{abstract}


\section{Introduction}
Quantum Machine Learning (QML) represents a rapidly evolving field at the intersection of quantum computing and artificial intelligence. By harnessing quantum phenomena such as superposition and entanglement, QML algorithms aim to outperform classical counterparts in complex tasks like optimization, pattern recognition, and generative modeling \cite{biamonte2017quantum, cerezo2022challenges, ciliberto2018quantum}. Among the various QML approaches, Hybrid Quantum Neural Networks (HQNNs) have emerged as powerful tools, demonstrating their potential in tackling a variety of complex challenges across domains such as medical image classification \cite{de2022survey,huang2023hybrid}, speech recognition \cite{thakar2024performance}, and cybersecurity threat detection \cite{suryotrisongko2022evaluating}. By leveraging the strengths of both quantum and classical computing, HQNNs offer enhanced computational efficiency and improved pattern recognition capabilities. The open-sourcing of quantum computing frameworks and QNNs has accelerated innovation, enabling researchers and developers to experiment with and deploy quantum-enhanced machine learning models on a large scale \cite{fingerhuth2018open}. However, this openness also introduces significant security vulnerabilities \cite{huang2023hybrid}. Publicly available implementations often lack robust protections against adversarial manipulations, making them susceptible to attacks that can compromise their integrity. While classical neural networks have been extensively studied for their susceptibility to adversarial threats, the security of HQNNs remains largely unexplored, particularly at the circuit level. This oversight presents a critical risk, as attackers can exploit architectural weaknesses in quantum circuits to degrade model performance without leaving detectable traces.

In this paper, we propose a novel circuit-level attack, which we call SQUASH, specifically targeting HQNNs used for classification tasks. Unlike conventional attack strategies that might focus on altering input data \cite{chu2023qdoor,john2025quantum}, SQUASH operates by directly manipulating the quantum circuit's structure through the insertion of SWAP gate(s) into the variational quantum circuit of the targeted HQNN. SWAP gates are fundamental quantum operations used to exchange the states of two qubits, facilitating proper qubit connectivity in quantum circuits, especially on hardware with limited qubit interactions. Their strategic insertion can misalign qubits, disrupting the intended evolution of quantum states and making them a potent tool for structural attacks like SQUASH.

Our proposed attack method is designed to be highly stealthy for several key reasons. Firstly, access to the original training data used to train the HQNN is not necessary. Secondly, SQUASH does not introduce any detectable perturbations in the input states that are fed into the network; instead, the disruption occurs within the quantum processing itself. We present two key attack strategies based on SWAP gate insertion. 1) Untargeted SWAP attacks aim to systematically degrade overall classification accuracy by perturbing fidelity measurements across all qubits (including the ancilla qubit). 2) Targeted SWAP attacks are designed to induce specific misclassifications, ensuring a given input is assigned to an adversary-controlled incorrect category by selectively manipulating fidelity measurements. This is done after correlating fidelity measurements to prediction outcomes and targeting specific qubits. In this two-pronged attack approach, SQUASH is able to reduce classification accuracy on average by 74.08\%  using our untargeted SWAP attack and 79.78\% using our targeted SWAP attack. These findings expose a critical vulnerability present in publicly available HQNN implementations and open-source quantum computing frameworks at the circuit level. 

The main contribution of this paper is summarized below:
\begin{enumerate}
    \item We propose a novel two-pronged attack strategy that utilizes SWAP gates to directly interfere with model performance using quantum entanglement.
    \item We highlight vulnerabilities within the HQNN architecture that have not been explored extensively in existing literature.
    \item By using the targeted SWAP attack strategy, we demonstrate the stealth of this attack by showing the classification performance becoming less distinguishable from normal behavior, with only subtle degradations in accuracy and misclassifications occurring in a targeted manner.
\end{enumerate}

\section{Related Work}
\subsection{Backdoor Attacks}  
Data-Poisoning-based Backdoor Attacks (DPBAs) and Trojaning attacks are common threats to neural networks, where adversaries manipulate training data or modify network components to embed hidden behaviors \cite{liu2018trojaning}. These attacks compromise model integrity, triggering malicious actions only under specific conditions. Related work \cite{kundu2024adversarial} addresses the unexplored threat of data poisoning attacks on QML models in cloud settings. They propose a novel quantum indiscriminate data poisoning attack, QUID, which leverages intra-class encoder state similarity to flip labels of training data, effectively degrading QML model performance. Through experiments, QUID demonstrates significant accuracy degradation, even in noisy environments and against classical defenses, highlighting the vulnerability of QML to data poisoning. QDoor is a backdoor attack framework for QNNs that differs from DPBAs by leveraging approximate synthesis \cite{chu2023qdoor}. It trains QNNs to behave normally until synthesis activates malicious behavior, exploiting unitary differences between uncompiled and synthesized models. QDoor enhances stealth, is robust against noise, and improves both attack success rate and clean data accuracy, making it harder to detect than traditional DPBAs.  

\subsection{Quantum Attacks}  
QTrojan is a circuit-level backdoor attack on QNNs that inserts quantum gates into the variational quantum circuit rather than modifying inputs or training data \cite{chu2023qtrojan}. It exploits quantum compiler configuration files as triggers, enabling stealthy attacks with improved clean data accuracy (21\%) and attack success rate (19.9\%). However, its effectiveness depends on users frequently updating configuration files and attackers successfully inserting triggers, limiting its reliability. Investigating SWAP tests for attacks has been investigated recently, with the authors proposing a SWAP-based fault injection model in multi-tenant quantum computing environments, targeting superconducting hardware systems \cite{upadhyay2023stealthyswapsadversarialswap}. The attack injects SWAP operations by strategically occupying qubits, increasing SWAP overhead by up to 55\% (median 25\%). They introduce a qubit quality metric for adversary selection and propose a machine learning model to detect anomalous user behavior as a countermeasure.  

In contrast, our work focuses on injecting SWAP gates into smaller scale models that are more accessible through open-source projects. The average user is more likely to interact with these smaller-scale, publicly available HQNN implementations rather than the more complex multi-tenant computing environments \cite{upadhyay2023stealthyswapsadversarialswap}. This makes the vulnerabilities exposed by SQUASH potentially more relevant and accessible to a wider range of users and developers experimenting with hybrid quantum machine learning \cite{reers2024comparative}.

\section{Background}

\subsection{Qubits and Quantum Circuits}
A qubit (quantum bit) is the basic unit of quantum information, representing a quantum analogue to the classical bit. Unlike a classical bit, which can only take the value 0 or 1, a qubit can exist in a superposition of both states simultaneously. The general state of a qubit is described as:
\begin{equation}
|\psi\rangle = \alpha|0\rangle + \beta|1\rangle,
\end{equation}
where $\alpha$ and $\beta$ are complex probability amplitudes such that $|\alpha|^2 + |\beta|^2 = 1$. These squared magnitudes represent the probability of measuring the qubit in the corresponding basis state.

The state of a qubit is often visualized using the Bloch sphere, where any pure state corresponds to a point on the surface of a unit sphere. This representation is useful for understanding how quantum gates transform qubit states through rotations and phase shifts. Furthermore, qubits can become entangled, meaning the state of one qubit cannot be described independently of the state of another. Entanglement is a uniquely quantum resource that enables powerful correlations and is critical for quantum algorithms and communication protocols \cite{biamonte2017quantum,nielsen2010quantum}. Qubit superposition and entanglement are essential for implementing the SWAP test, which relies on coherent interference between qubit states to measure their similarity.

\begin{figure*}[!t]
    \centering
    \includegraphics[width=0.8\textwidth]{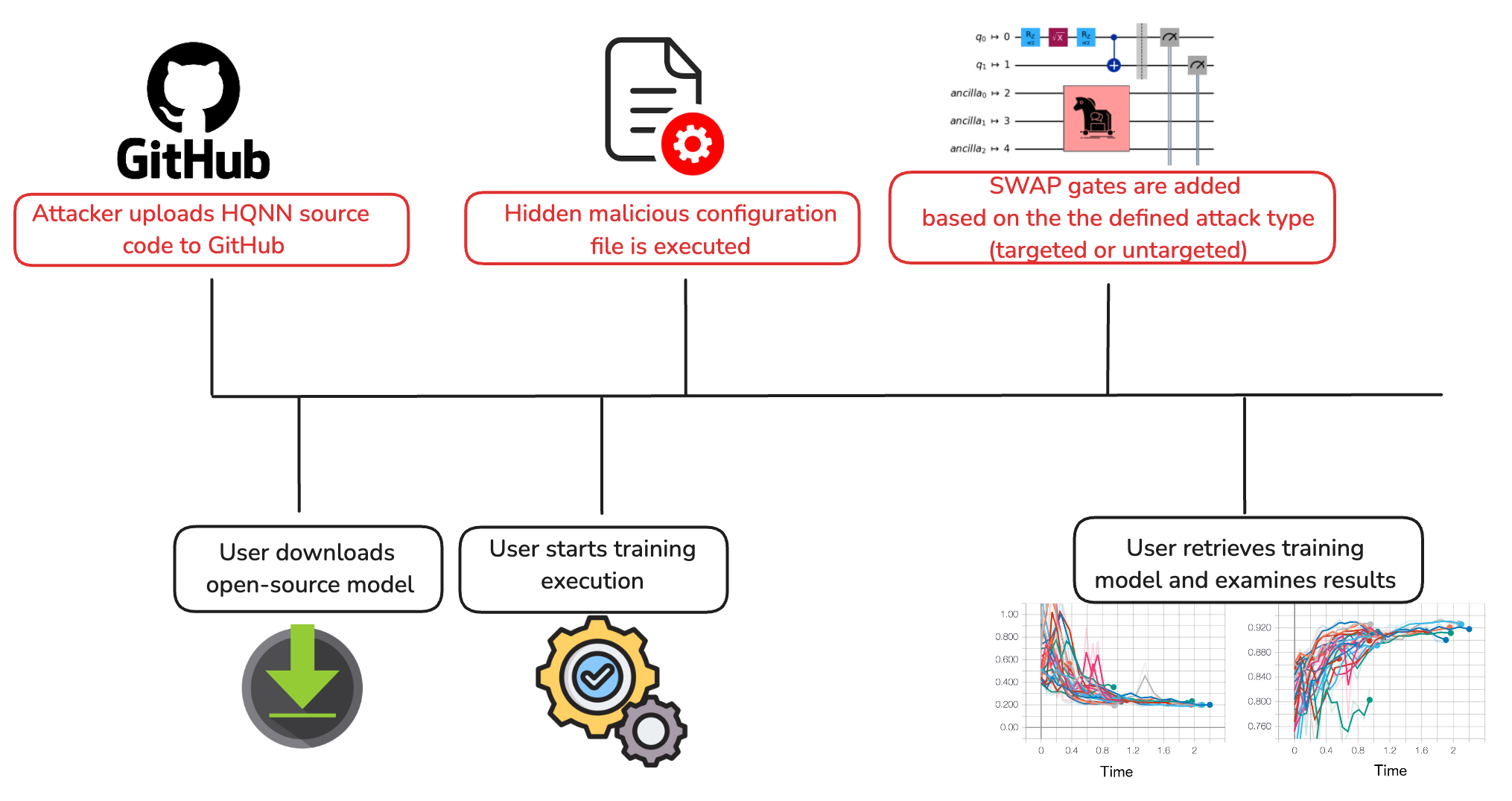}
    \caption{Illustration of the user sequence that results in SQUASH being successful. After a user has successfully downloaded the attacker's HQNN code, the configuration file injects SWAP tests (based on the type of attack) after the user has initiated the training execution phase.}
    \label{fig:context}
\end{figure*}

To manipulate qubits, quantum computers use quantum circuits, which are composed of quantum gates. These gates are unitary operators that evolve the state of qubits in a reversible and deterministic manner \cite{schuld2020circuit}. For a single-qubit gate $U$, the transformation is:
\begin{equation}
|\psi\rangle \rightarrow U|\psi\rangle.
\end{equation}
Single-qubit gates act on individual qubits and perform basic operations. For example, the Pauli-X gate behaves like a classical NOT gate by flipping $|0\rangle$ to $|1\rangle$ and vice versa. The Hadamard gate (H) puts a qubit into an equal superposition of $|0\rangle$ and $|1\rangle$, while the phase gates (S and T) apply phase shifts.

Multi-qubit gates, such as the Controlled-NOT (CNOT) gate, are essential for enabling interactions and entanglement between qubits. A CNOT gate flips the target qubit if the control qubit is in state $|1\rangle$. These gates are the backbone of conditional operations and entanglement generation in quantum circuits. A quantum circuit is constructed by applying a series of single- and multi-qubit gates to an initial state (typically $|0\rangle^{\otimes n}$ for $n$ qubits). As the gates are applied, the qubit states evolve to encode the solution of a computational task.

\subsection{SWAP Tests}

The SWAP test is a fundamental quantum algorithm used to determine the similarity between two quantum states, making it particularly valuable in HQNNs for tasks such as quantum kernel estimation and entanglement detection  \cite{foulds2021controlled,liu2024quantum, maldonado2024quantum}. The test operates by introducing an ancilla qubit and applying a controlled-SWAP (CSWAP) gate to swap between \textit{N} target states, followed by a Hadamard transformation on the ancilla qubit \cite{kay2018tutorial}. Mathematically, given two quantum states $|\psi\rangle$ and $|\phi\rangle$, and an ancilla initialized as $|0\rangle$, the total system is prepared as:
\begin{equation}
    |\Psi\rangle = |0\rangle \otimes |\psi\rangle \otimes |\phi\rangle.
\end{equation}Applying a Hadamard gate to the ancilla results in:\begin{equation}
    \frac{1}{\sqrt{2}} (|0\rangle + |1\rangle) \otimes |\psi\rangle \otimes |\phi\rangle.
\end{equation}
The CSWAP gate is then applied, which swaps $|\psi\rangle$ and $|\phi\rangle$ if the ancilla is in state $|1\rangle$, yielding:\begin{equation}
    \frac{1}{\sqrt{2}} \left( |0\rangle \otimes |\psi\rangle \otimes |\phi\rangle + |1\rangle \otimes |\phi\rangle \otimes |\psi\rangle \right).
\end{equation}
A final Hadamard transformation on the ancilla leads to a probability of measuring $|0\rangle$ given by:
\begin{equation}
    P(0) = \frac{1 + |\langle \psi | \phi \rangle|^2}{2}.
\end{equation}

This probability directly encodes the squared fidelity between the two states, making the SWAP test an essential tool in HQNNs for assessing quantum state similarity in classification and clustering tasks. However, the SWAP test is vulnerable to noise and decoherence, especially in near-term quantum devices where gate errors impact fidelity estimation \cite{ripper2023swap}. As the number of qubits increases, maintaining coherence becomes more challenging due to the exponentially larger Hilbert space and increased potential for qubit interactions that induce decoherence. This noise can distort quantum operations, including the SWAP test, by introducing gate errors that affect superposition states and lead to inaccurate fidelity measurements. These inaccuracies compromise the SWAP test’s ability to measure quantum state overlaps, making it susceptible to exploitation in quantum attacks where adversaries extract information about unknown states by measuring their overlap with known references \cite{liu2024quantum}.

\section{SQUASH}

\subsection{Overview of SQUASH}
We propose SQUASH as a circuit-level attack to sabotage HQNNs designed for classification tasks. More specifically, we achieve this by inserting SWAP gate(s) directly into the variational quantum circuit of the targeted HQNN. Unlike conventional attacks that might introduce noise or manipulate input data, SQUASH operates by directly altering the quantum circuit's structure, leading to qubit misalignment and consequently disrupting performance. This method of attack is considered highly stealthy for several reasons. Firstly, access to the original training data used to train the HQNN is not necessary. Secondly, SQUASH does not introduce any detectable perturbations in the input states that are fed into the network. Instead, the disruption occurs within the quantum processing itself.

For these experiments, we assume that the user are in need of an open-sourced QNN model for a classification task. Given the model efficiency of utilizing quantum and classical computations, the user finds an implementation of an HQNN on GitHub and decides to download it for their task. The user, unaware of the potential risks, downloads the model directly from a public repository. While the model itself appears to be open-sourced and reputable, the repository also includes a malicious configuration file that is designed to exploit vulnerabilities within the user's system \cite{john2025quantum, chu2023qtrojan}. This configuration file is carefully crafted to inject harmful code when the user attempts to load the model into their environment. Once the HQNN model is downloaded and the configuration file is executed, the malicious code triggers specific actions specified by the attacker, such as degrading model performance or intentionally mispredicting specific classes. The user unknowingly opens the door to a security breach simply by downloading the model from GitHub without properly reviewing the associated files. The complete flow is shown in Figure \ref{fig:context}.

\begin{figure}[!t]
\begin{minipage}[t]{0.45\textwidth}
\centering
\begin{algorithm}[H]
\caption{\textbf{SWAP Attack for HQNN}}
\label{alg:1}
\begin{algorithmic}[1]
\State \textcolor{blue}{/* Step 1: Initialize parameters */}
\State Set initial perturbation $\delta \gets 0$
\For{each qubit $\psi$ in the quantum circuit}
    \State \textcolor{blue}{/* Step 2: Check attack type */}
    \If{attack\_type == untargeted}
        \State Apply random SWAP gates between circuit qubits
    \Else
    \State Compute initial fidelities:
    \Statex \quad $F_t \gets F(|\psi_t\rangle, |x\rangle)$
    \Statex \quad $F_c \gets F(|\psi_c\rangle, |x\rangle)$
    
    \State \textcolor{blue}{/* Step 3: Optimize perturbation for qubit $\psi_i$ */}
        \While{$F_t \leq F_c$}
            \State Compute gradient $\nabla_{\delta} F(|\psi_t\rangle, |x + \delta \rangle)$
            \State Update perturbation: $\delta \gets \delta + \eta \cdot \nabla_{\delta}$
            \State Generate perturbed state: $|x'\rangle \gets e^{i\delta} |x\rangle$
            \State Recalculate fidelities:
            \Statex \quad $F_t \gets F(|\psi_t\rangle, |x'\rangle)$
            \Statex \quad $F_c \gets F(|\psi_c\rangle, |x'\rangle)$
        \EndWhile
    \EndIf
\EndFor
\State \textcolor{blue}{/* Step 4: Return adversarial state */}
\State \textbf{return} $|\psi_t\rangle$
\end{algorithmic}
\end{algorithm}
\end{minipage}

\end{figure}

\subsection{Adversarial SWAP-Based Attacks on Hybrid Quantum Neural Networks}

Hybrid quantum-classical neural networks leverage quantum fidelity measurements as a fundamental component of their classification process \cite{el2024advqunn,el2024robqunns}. Specifically, the SWAP test serves as a mechanism to evaluate the similarity between quantum states, enabling the construction of class boundaries based on quantum feature representations \cite{ripper2023swap}. However, this reliance introduces vulnerabilities that adversaries can exploit to degrade classification accuracy or induce targeted misclassification. In this work, we utilize SWAP tests to alter classification performance by leveraging measured probabilities to reinforce quantum state grouping within HQNNs. Our approach consists of two key attack strategies: untargeted SWAP attacks and targeted SWAP attacks, both designed to manipulate classification outcomes by influencing quantum fidelity measurements. The algorithm for inserting the SWAP tests into the circuit is outlined in Algorithm \ref{alg:1}.

\subsubsection{Untargeted SWAP Attacks}

\begin{figure}[!t]
    \centering
    \begin{quantikz}
    \lstick{\ket{0}} & \gate{H} & \ctrl{2} & \ctrl{1} \gategroup[3,steps=3,style={dashed,rounded
    corners,fill=red!20, inner
    xsep=2pt},background,label style={label
    position=below,anchor=north,yshift=-0.2cm}]{{\sc
    malicious swap}} & \targX{} & \targX{} & \gate{H} & \meter{} \\
    \lstick{\ket{\chi}} & & \swap{1} & \targX{} & \ctrl{-1} & \targX{} & &  & \qw \\
    \lstick{\ket{\psi}} & & \targX{} & \qw & \ctrl{-1} & \ctrl{-2}& \qw & \qw & \qw
    \end{quantikz}
    \caption{An untargeted attack on a 2-qubit quantum circuit. The first connection between \ket{0} with the crosses on \ket{\chi} and \ket{\psi} represents a healthy SWAP test. The connections in the MALICIOUS SWAP container represent SWAP tests that contaminate the fidelities of \ket{\chi} and \ket{\psi} due to improper configuration.}
    \label{fig:untargeted_attack_circuit}
\end{figure}
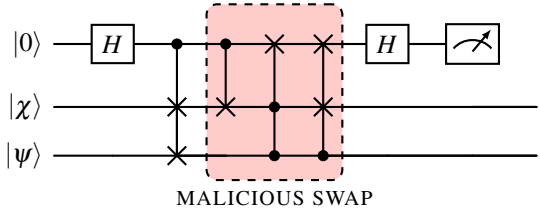

An untargeted SWAP attack aims to systematically degrade classification accuracy by covertly injecting multiple SWAP gates across all qubits, as shown in Figure \ref{fig:untargeted_attack_circuit}. Given an input qubit state \( |x\rangle \), the attack disrupts the similarity computations essential for accurate label assignment for classification. The similarity metric is given by:
\begin{equation}
    S_i = \sum_{j} P_i^{(j)}(0) = \sum_{j} \frac{1 + |\langle x | r_j^{(i)} \rangle|^2}{2},
\end{equation}
where \( P_i^{(j)}(0) \) is the probability obtained from the SWAP test between \( |x\rangle \) and the reference state \( |r_j^{(i)}\rangle \). The classifier assigns \( |x\rangle \) to the category maximizing \( S_i \), ensuring it is grouped with the most similar reference states.

By injecting SWAP gates indiscriminately across all qubits—including ancilla qubits—the adversary introduces subtle interference that degrades fidelity across multiple class comparisons. Unlike targeted attacks, which aim to misclassify specific inputs, this broad perturbation reduces overall confidence in predictions, leading to increased classification uncertainty. Repetitive SWAP tests cause discrepencies with fidelities of qubit measurements since the entanglement between qubits can be disrupted, leading to a loss of correlation that results in discrepancies across fidelities. Since quantum classification is often embedded within a hybrid learning framework, these perturbations propagate through subsequent layers, distorting extracted features and degrading the overall performance of the model. Additionally, as hybrid models employ iterative refinement strategies for ambiguous classifications, an untargeted SWAP attack can amplify model instability, forcing redundant evaluations and increasing computational overhead.

\subsubsection{Targeted SWAP Attacks}  

In contrast to untargeted attacks, a targeted SWAP attack is designed to induce a specific misclassification, ensuring that a given input state is systematically assigned to an incorrect but adversary-controlled category. The attack selectively manipulates the fidelity function to ensure:
\begin{equation}
    F(|\psi_t\rangle, |\chi\rangle) > F(|\psi_c\rangle, |\chi\rangle),
\end{equation}
where \( |\psi_t\rangle \) represents a reference state from the adversary-chosen target class, and \( |\psi_c\rangle \) corresponds to the reference state from the correct class as shown in Figure \ref{fig:targeted_attack_circuit}. By artificially increasing the measured fidelity between \( |\chi\rangle \) and \( |\psi_t\rangle \), while simultaneously reducing its similarity to \( |\psi_c\rangle \), the adversary forces the classifier to misattribute the input to the incorrect category as described in Algorithm \ref{alg:1}. Targeted SWAP attacks in SQUASH only use two SWAP tests to carry out the attack, as the qubit fidelities are measured using SWAP before and after the circuit tampering, hence reducing the overhead of the circuit execution on the user's machine.   

\tikzset{
noisy/.style={starburst,fill=yellow,draw=red,line
width=2pt,inner xsep=-4pt,inner ysep=-5pt}
}


\begin{table*}[!t]
    \centering
    \begin{tabular}{llcccc}
        \toprule
        Dataset & SWAP Count & \multicolumn{2}{c}{Clean Training} & \multicolumn{2}{c}{Untargeted SWAP Attack} \\
        \cmidrule(lr){3-4} \cmidrule(lr){5-6}
                 &         & NLL & Accuracy (\%) & NLL & Accuracy (\%) \\
        \midrule
        \multirow{3}{*}{MNIST-2} 
        & 1 SWAP Test & 0.0451 & 96.01\% & 0.512 & 43.91\% \\
        & 2 SWAP Tests & 0.0510 & 95.68\% & 0.649 & 35.22\% \\
        & 3 SWAP Tests & 0.0687 & 95.30\% & \textbf{0.778} & \textbf{21.78\%} \\
        \midrule
        \multirow{3}{*}{CIFAR-2} 
        & 1 SWAP Test & 0.1102 & 90.78\% & 0.599 & 38.04\% \\
        & 2 SWAP Tests & 0.1127 & 91.41\% & 0.769 & 30.92\% \\
        & 3 SWAP Tests & 0.1210 & 91.87\% & \textbf{0.932} & \textbf{18.12\%} \\
        \midrule
        \multirow{3}{*}{MNIST-10} 
        & 1 SWAP Test & 0.0991 & 93.11\% & 0.598 & 38.27\% \\
        & 2 SWAP Tests & 0.0904 & 93.88\% & 0.701 & 32.64\% \\
        & 3 SWAP Tests & 0.0904 & 93.89\% & \textbf{0.880} & \textbf{19.31\%} \\
        \midrule
        \multirow{3}{*}{CIFAR-10} 
        & 1 SWAP Test & 0.1723 & 77.42\% & 0.651 & 34.07\% \\
        & 2 SWAP Tests & 0.1771 & 78.36\% & 0.744 & 27.18\% \\
        & 3 SWAP Tests & 0.1817 & 78.92\% & \textbf{0.911} & \textbf{15.84\%} \\
        \bottomrule
    \end{tabular}
    \caption{Classification results across MNIST and CIFAR using the 2-qubit HQNN under clean training and untargeted SWAP attack conditions.}
    \label{tab:multi_dataset_swap_attack}
    \vspace{-5pt}
\end{table*}

\subsection{Circuit Design}
This architecture combines classical convolutional layers with a QNN to leverage both classical and quantum computing for classification tasks. It begins with two convolutional layers that process input data through feature extraction, detecting patterns and spatial hierarchies in the data. A dropout layer is included for regularization, preventing overfitting by randomly deactivating a fraction of neurons during training.

After the convolutional layers, the output is flattened and passed through two fully connected layers, which reduce the dimensionality and prepare the data for input into the quantum network. The second fully connected layer outputs a 2-dimensional tensor, which is then passed to the quantum network. The quantum processing begins with a feature map layer, which is responsible for encoding the classically processed data into the quantum states of the network's qubits. The number of qubits used in this encoding layer is 2 qubits. Following the encoding layer is a variational ansatz, which consists of parameterized quantum blocks. These blocks apply a sequence of adjustable quantum gates to the encoded quantum state, allowing the network to learn complex relationships in the data. The quantum output is then passed through a final fully connected layer, producing a 1-dimensional output that is normalized to form a 2-dimensional result, which is concatenated as the final classification output.

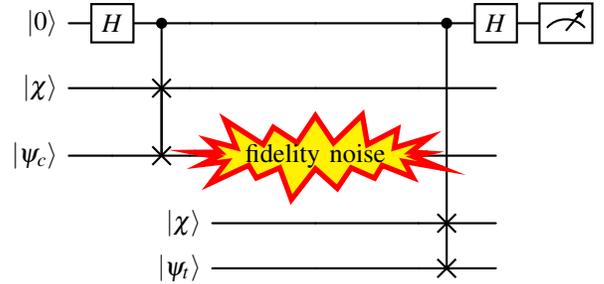
\begin{figure}[!t]
    \centering
    \begin{quantikz}[row sep=0.6cm,column sep=0.3cm,wire
    types={q,q,q,n,n}]%
    \lstick{\ket{0}} & \gate{H} & \ctrl{2} & &  &  &  & \ctrl{4} & \gate{H} & \meter{}\\
    \lstick{\ket{\chi}} & & \swap{1} & & & &  & & \\
    \lstick{\ket{\psi_c}} & & \targX{} & & & \gate[1,style={noisy},label style=black]{\text{fidelity noise}} & &  & \\
    &&&& \lstick{\ket{\chi}} & \setwiretype{q} & 
    & \targX{} &  \\
    &&&& \lstick{\ket{\psi_t}} & \setwiretype{q} & &
    \targX{} & 
    \vphantom{\lstick{}} 
    \end{quantikz}
    \caption{A targeted attack on a 2-qubit quantum circuit. \ket{\psi_c} represents the correct classification and \ket{\psi_t} represents the attack target classification.}
    \label{fig:targeted_attack_circuit}
\end{figure}

A crucial and distinctive element of this HQNN architecture is the inclusion of a swap test performed on a specific pair of qubits. This SWAP Test serves as the measurement layer in this context, as it determines the fidelity (similarity) between quantum states, which is then used for classification. The swap test involves an ancilla qubit, Hadamard gates, and a CSWAP gate. The probability of measuring the ancilla qubit in a certain state provides a measure of the fidelity. Finally, a fully connected layer converts the qubit outputs and fidelities to an MNIST classification.

\section{Experimental Results}

\subsection{Experimental Setup}
We utilize Qiskit, IBM’s open-source quantum software development kit, in conjunction with a custom Python wrapper to facilitate quantum circuit simulations. To assess the performance of the HQNN across different data complexities, we train the model on both the MNIST and CIFAR-10 datasets. MNIST serves as a standard benchmark for handwritten digit recognition, consisting of 70,000 grayscale images of digits (0–9) with a resolution of 28×28 pixels \cite{lecun1998gradient}. CIFAR-10, in contrast, comprises 60,000 RGB images of size 32×32 pixels, spanning 10 classes that include vehicles and animals \cite{krizhevsky2014cifar}. To evaluate performance under different classification complexities, we extract both binary and multiclass subsets from each dataset. For MNIST-2, we select digits 0 and 1 to form a binary classification task with minimal intra-class variation. For CIFAR-2, we use the cat and dog classes, representing a more visually complex binary problem due to overlapping textures and poses. These binary tasks help highlight the HQNN’s sensitivity to visual abstraction in low- and high-complexity domains. The full multiclass tasks (MNIST-10 and CIFAR-10) retain all 10 classes from each dataset, offering a broader perspective on generalization across diverse categories. In all cases, class-balanced splits are maintained between training and testing to ensure fair evaluation. Both datasets are preprocessed using PyTorch’s \textit{torchvision} library \cite{paszke2019pytorch}, transformed into tensors via \textit{torchvision.transforms}, and loaded using PyTorch’s \textit{DataLoader}, with a fixed random seed applied to ensure reproducibility during training.

\subsection{Evaluation Metrics}
Negative Log Loss (NLL) is a widely used metric for evaluating classification models, particularly in HQNNs. This metric is suitable for HQNN-based image classification because quantum circuits inherently produce probabilistic outputs. The measurement results from a PQC can be interpreted as probability distributions over class labels, making NLL a natural choice for assessing model confidence. In an image classification task with \( C \) possible classes, let \( p(y_c \mid \mathbf{x}) \) represent the predicted probability of class \( c \) for an input image \( \mathbf{x} \). The NLL for a single sample is calculated as:
\begin{equation}
\mathcal{L} = -\sum_{c=1}^{C} y_c \log p(y_c \mid \mathbf{x}),
\end{equation}
where \( y_c \) is a binary indicator (1 for the correct class, 0 otherwise). The total loss across a dataset with \( N \) samples is given by:
\begin{equation}
\mathcal{L}^N = -\frac{1}{N} \sum_{i=1}^{N} \sum_{c=1}^{C} y_{i,c} \log p(y_{i,c} \mid \mathbf{x}_i),
\end{equation}
where \( y_{i,c} \) and \( p(y_{i,c} \mid \mathbf{x}_i) \) represent the ground truth and predicted probability for sample \( i \), respectively.

NLL is particularly useful for HQNNs due to its compatibility with probabilistic quantum measurements. Quantum circuits produce output distributions according to the Born rule, where probabilities are obtained from quantum state amplitudes \cite{neumaier2019born}. Since NLL penalizes incorrect classifications more heavily when assigned high confidence, it effectively guides HQNNs to refine their probability estimates.

\begin{table*}[!t]
    \begin{centering} 
    \begin{tabular}{lccccc}
        \toprule
        \multirow{2}{*}{} & \multicolumn{2}{c}{Clean Training} & \multicolumn{3}{c}{Targeted SWAP Attack} \\
        \cmidrule(lr){2-3} \cmidrule(lr){4-6} 
                          & NLL & Total Accuracy (\%)  & NLL & Target Class Accuracy (\%) & Total Accuracy (\%)\\
        \midrule
        MNIST-2 & 0.0510 & 95.68\% & 0.823 & 19.14\% & 57.075\%\\
        \midrule
        CIFAR-2 & 0.1123 & 91.38\% & 0.932 & 18.12\% & 54.062\%\\
        \midrule
        MNIST-10 & 0.952 & 92.3\% & 0.801 & 20.94\% & 85.164\%\\
        \midrule
        CIFAR-10 & 0.1817 & 78.92\% & 0.911 & 15.84\% & 73.919\%\\
        \bottomrule
    \end{tabular}
    \caption{Results for classification of MNIST and CIFAR datasets (2 and 10 classes) under clean training and targeted SWAP attack. Target Class Accuracy measures performance specifically on the attacker’s chosen class, while Total Accuracy reflects overall model performance. A successful attack lowers both, with a more severe drop in the target class.}

    \label{tab:targeted_swap_attack}
    \end{centering}
    \vspace{-15pt}
\end{table*}

\subsection{Classification Analysis}
The performance of the HQNN under clean training conditions—absent any adversarial SWAP interference—is summarized in Table \ref{tab:multi_dataset_swap_attack}. Across all datasets, the model demonstrated strong classification capability, with accuracy values consistently high and NLL values low, even as the number of inserted SWAP tests increased. On simpler datasets such as MNIST-2, accuracy remained above 95\%, while CIFAR-2 yielded slightly lower but still robust results around 91\%. For the more complex multiclass tasks, MNIST-10 maintained performance above 93\%, and CIFAR-10 achieved accuracy near 79\%. Interestingly, for CIFAR datasets, the accuracy slightly increased with more SWAP tests under clean training. This may reflect a beneficial effect of additional circuit entanglement when processing richer, high-dimensional data, in contrast to MNIST where performance stays consistent due to the simplicity of the input space.

\begin{figure}[!t]
    \centering
    \includegraphics[width=0.45\textwidth]{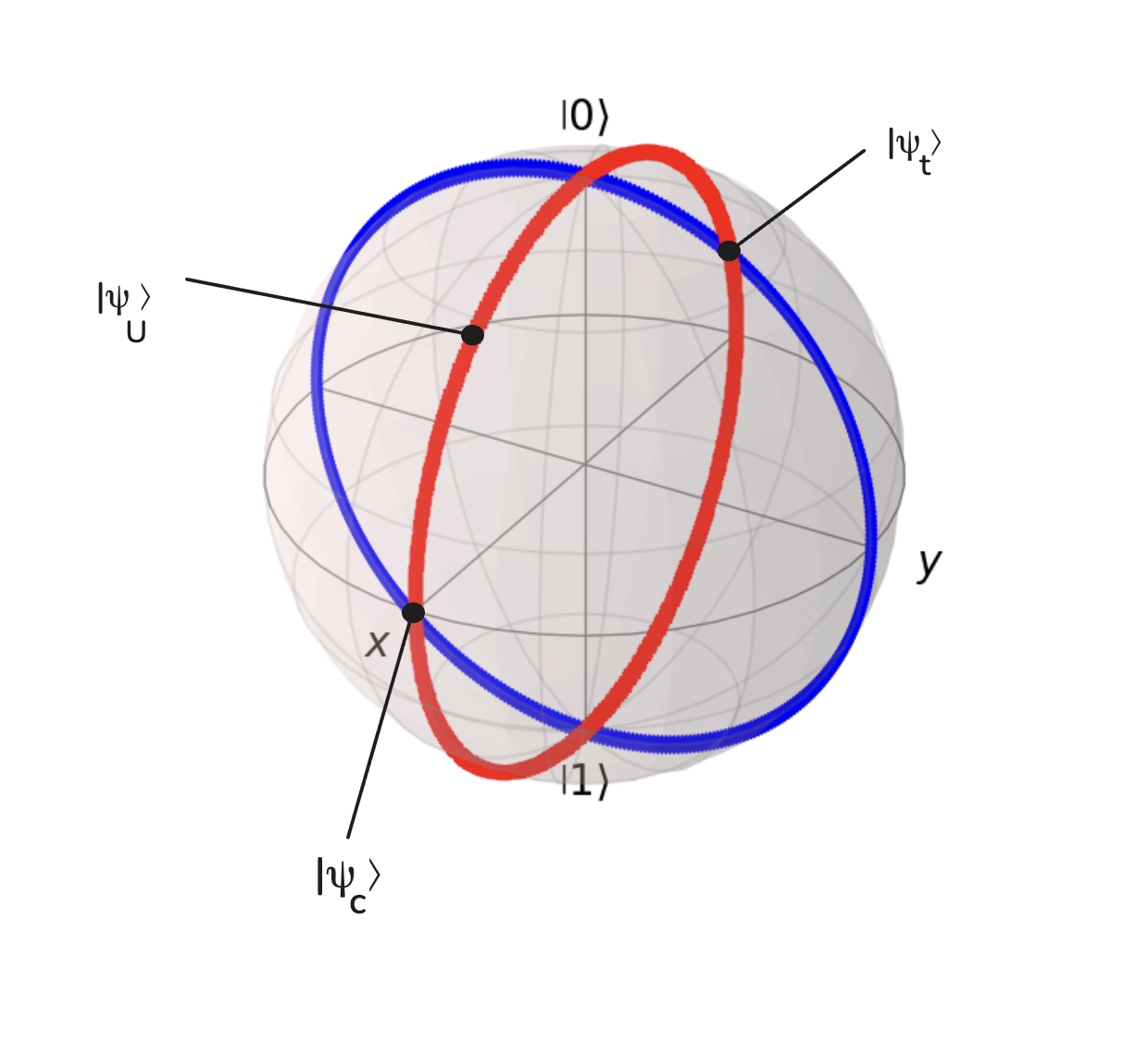}
    \caption{Visualization of a Bloch sphere with an untargeted and targeted SQUASH attack. The blue ring represents a clean Hadamard gate, whereas the red ring represents a Hadamard gate from a SWAP attack. The qubits represent the correct class from clean training (\( |\psi_c\rangle \)), the targeted class from the targeted attack (\( |\psi_t\rangle \)), and the randomized initialization from the untargeted attack (\( |\psi_u\rangle \)).}
    \label{fig:bloch_sphere}
\end{figure}

Upon the introduction of untargeted SWAP attacks, a pronounced vulnerability in the HQNN framework was revealed. Across all datasets, increasing the number of maliciously inserted SWAP tests resulted in a consistent degradation of model performance, evidenced by declining classification accuracy and rising negative log-likelihood (NLL). For instance, on the MNIST-2 dataset, accuracy dropped from 96.01\% under clean training to 43.91\%, 35.22\%, and 21.78\% as one, two, and three SWAP tests were applied, respectively. A similar pattern was observed on CIFAR-2, where clean accuracy of 91.87\% fell sharply to 38.04\%, 30.92\%, and 18.12\% under increasing SWAP interference. The multiclass settings, MNIST-10 and CIFAR-10, also exhibited substantial degradation: MNIST-10 accuracy declined from 93.89\% to 19.31\%, while CIFAR-10 dropped from 78.92\% to just 15.84\%. Concurrently, NLL values rose across all scenarios, indicating increasing model uncertainty. These results reinforce the hypothesis that untargeted SWAP attacks broadly perturb the fidelity measurements by inducing structural misalignments in the quantum circuit, as shown in Figure \ref{fig:bloch_sphere}. Here, \( |\psi_u\rangle \) is the qubit with randomized fidelity on the faulty Hadamard gate, where its arbitrary location leads to severely degraded classification performance for classes relying on that specific qubit.

The targeted SWAP attacks exhibited a distinct pattern, specifically designed to manipulate the accuracy of a single target class as shown in Table \ref{tab:targeted_swap_attack}. For the MNIST-2 dataset, the accuracy for the target class dropped significantly to 19.14\% under the targeted SWAP attack. Similar trends were observed for MNIST-10, where target class accuracy reduced to 20.94\%. These drastic drops in accuracy demonstrate the attack’s effectiveness in inducing targeted misclassifications by manipulating fidelity measurements for the chosen class, as illustrated by \( |\psi_t\rangle \) in Figure~\ref{fig:bloch_sphere}. In this figure, \( |\psi_t\rangle \) lies at the intersection of the clean Hadamard gate (blue) and the Hadamard gate from the SWAP attack circuit (red). This overlap conceals the qubit’s altered state by making it appear consistent with the clean circuit, thereby rendering the attack stealthy at the quantum level.
Despite the significant drop in target class accuracy, the overall accuracy showed a more moderate decline when compared to clean training. For instance, for MNIST-2, the total accuracy under a targeted SWAP attack decreased to 57.075\%, a considerable drop from the clean training accuracy of 95.01\%. Similarly, for MNIST-10, the total accuracy dropped to 85.164\%, compared to the clean training accuracy of 92.3\%. This is a result of the correct classifications of other classes masking the incorrect classification of the target class, making the attack stealthy to the end user. These results indicate that while the attack is highly effective at targeting and misclassifying a specific class, the network’s overall ability to classify other instances remains largely intact.

\subsection{SWAP Overhead Analysis}
\begin{figure}[!t]
\centering
    \includegraphics[width=0.85\columnwidth]{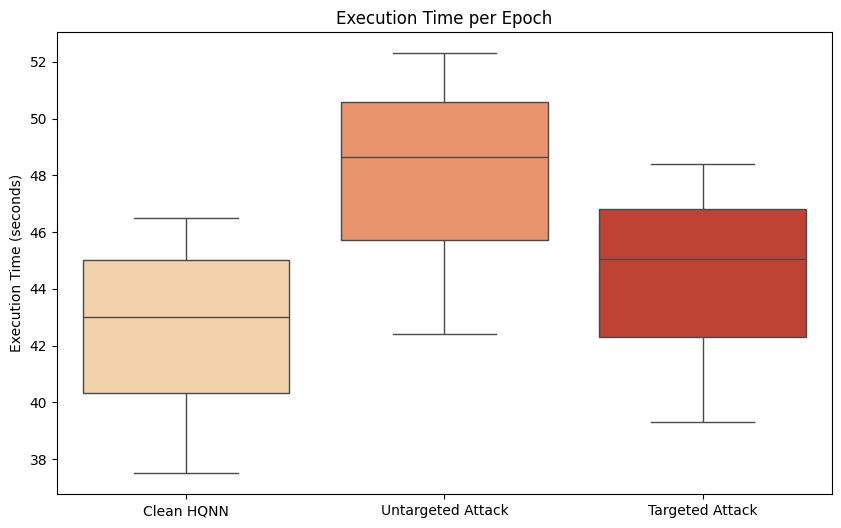}
    \caption{Visualization of the execution times across 100 executions for each model.}
    \label{fig:swap_overhead}
    \vspace{-5pt}
\end{figure}

To evaluate the overhead of conducting SWAP operations through the SQUASH attack, we measure the SWAP overhead as the percentage increase in execution time per epoch for both MNIST and CIFAR combined. In our experiments, we conducted 100 training iterations for each model and recorded the execution time for each epoch. From Figure \ref{fig:swap_overhead}, the targeted SWAP attack model exhibited an average execution time of 44.38 seconds, corresponding to a 5.96\% increase compared to the clean HQNN model, which averaged 42.07 seconds. The untargeted SWAP attack model demonstrated a significantly higher average execution time of 48.82 seconds, reflecting a 16.05\% increase over the clean HQNN model. The increased execution time for both attack models can be attributed primarily to the additional CNOT gates required by the SWAP gate decomposition (typically involving three CNOT gates per SWAP). This, in turn, increases the circuit depth and the number of operations necessary during training. Furthermore, due to the inherent nature of the SQUASH attack, untargeted attacks require more SWAP tests to degrade classification performance, whereas targeted attacks strategically add fidelity only utilizing two SWAP tests. As a result, the overhead associated with untargeted attacks is more pronounced.

\subsection{Potential Defenses}
Given that SQUASH operates by directly inserting SWAP gates to induce qubit misalignment and disrupt quantum state evolution, monitoring quantum circuit structure and execution for anomalous activity represents a key defensive strategy. This could involve the development of anomaly detection systems capable of identifying unexpected increases in the number of SWAP gates or deviations from the intended circuit architecture during the execution phase \cite{upadhyay2023stealthyswapsadversarialswap, ghosh2024ai}. Solutions like QML-IDS aim to detect attacks before hitting the circuit layer, which can potentially pose as a problem for attackers wanting to use SQUASH \cite{abreu2024qml}. Furthermore, the exploitation of publicly available QML implementations through malicious configuration files underscores the need for enhanced security measures concerning the provenance and validation of quantum code and associated configuration files before deployment and training \cite{li2016survey}. Ultimately, addressing the circuit-level vulnerabilities highlighted by SQUASH requires the development and integration of robust defense mechanisms specifically designed to detect and prevent unauthorized modifications to the quantum components.

\section{Conclusion}
The presence of SWAP-based adversarial vulnerabilities highlights several critical security challenges in hybrid quantum-classical learning architectures. Perturbations to quantum fidelity scores directly impact classification reliability, increasing model uncertainty and reducing overall predictive accuracy. As fidelity-based decisions propagate through hybrid layers, errors introduced at the quantum stage distort feature representations, exacerbating misclassification risks. Furthermore, adaptive learning mechanisms that iteratively refine classification boundaries may reinforce adversarial misclassifications over time, further degrading model robustness.  

The stark contrast in the impact of untargeted and targeted SWAP attacks underscores the nuanced vulnerabilities present in HQNNs at the circuit level. The significant performance degradation under untargeted attacks, even with minimal circuit modifications, highlights a critical security gap in publicly available HQNN implementations and open-source quantum computing frameworks. This vulnerability stems from the reliance of HQNNs on quantum fidelity measurements for classification. The insertion of SWAP gates, as implemented in the SQUASH attack, proves to be a stealthy and effective method for sabotaging HQNNs without necessitating access to sensitive training data or introducing detectable input perturbations. The fact that increasing the number of inserted SWAP tests generally amplifies the detrimental effects of untargeted attacks further emphasizes the sensitivity of these hybrid architectures to seemingly minor structural alterations in their quantum components. The findings of this work suggest that while hybrid quantum-classical architectures leverage the strengths of both classical neural networks and QNNs to enhance high-dimensional feature representation, their reliance on fidelity-based classification creates vulnerabilities that can be exploited.

\bibliographystyle{IEEEtran}
\bibliography{IEEEabrv,references}

\begin{thebibliography}{10}
\providecommand{\url}[1]{#1}
\csname url@samestyle\endcsname
\providecommand{\newblock}{\relax}
\providecommand{\bibinfo}[2]{#2}
\providecommand{\BIBentrySTDinterwordspacing}{\spaceskip=0pt\relax}
\providecommand{\BIBentryALTinterwordstretchfactor}{4}
\providecommand{\BIBentryALTinterwordspacing}{\spaceskip=\fontdimen2\font plus
\BIBentryALTinterwordstretchfactor\fontdimen3\font minus \fontdimen4\font\relax}
\providecommand{\BIBforeignlanguage}[2]{{%
\expandafter\ifx\csname l@#1\endcsname\relax
\typeout{** WARNING: IEEEtran.bst: No hyphenation pattern has been}%
\typeout{** loaded for the language `#1'. Using the pattern for}%
\typeout{** the default language instead.}%
\else
\language=\csname l@#1\endcsname
\fi
#2}}
\providecommand{\BIBdecl}{\relax}
\BIBdecl

\bibitem{biamonte2017quantum}
J.~Biamonte, P.~Wittek, N.~Pancotti, P.~Rebentrost, N.~Wiebe, and S.~Lloyd, ``Quantum machine learning,'' \emph{Nature}, vol. 549, no. 7671, pp. 195--202, 2017.

\bibitem{cerezo2022challenges}
M.~Cerezo, G.~Verdon, H.-Y. Huang, L.~Cincio, and P.~J. Coles, ``Challenges and opportunities in quantum machine learning,'' \emph{Nature computational science}, vol.~2, no.~9, pp. 567--576, 2022.

\bibitem{ciliberto2018quantum}
C.~Ciliberto, M.~Herbster, A.~D. Ialongo, M.~Pontil, A.~Rocchetto, S.~Severini, and L.~Wossnig, ``Quantum machine learning: a classical perspective,'' \emph{Proceedings of the Royal Society A: Mathematical, Physical and Engineering Sciences}, vol. 474, no. 2209, p. 20170551, 2018.

\bibitem{de2022survey}
G.~De~Luca, ``A survey of nisq era hybrid quantum-classical machine learning research,'' \emph{Journal of Artificial Intelligence and Technology}, vol.~2, no.~1, pp. 9--15, 2022.

\bibitem{huang2023hybrid}
S.~Huang, Y.~Chang, Y.~Lin, and S.~Zhang, ``Hybrid quantum--classical convolutional neural networks with privacy quantum computing,'' \emph{Quantum Science and Technology}, vol.~8, no.~2, p. 025015, 2023.

\bibitem{thakar2024performance}
Y.~Thakar, B.~Ghosh, V.~Adeshra, and K.~Srivastava, ``Performance analysis of hybrid quantum-classical convolutional neural networks for audio classification,'' in \emph{2024 15th International Conference on Computing Communication and Networking Technologies (ICCCNT)}.\hskip 1em plus 0.5em minus 0.4em\relax IEEE, 2024, pp. 1--7.

\bibitem{suryotrisongko2022evaluating}
H.~Suryotrisongko and Y.~Musashi, ``Evaluating hybrid quantum-classical deep learning for cybersecurity botnet dga detection,'' \emph{Procedia Computer Science}, vol. 197, pp. 223--229, 2022.

\bibitem{fingerhuth2018open}
M.~Fingerhuth, T.~Babej, and P.~Wittek, ``Open source software in quantum computing,'' \emph{PloS one}, vol.~13, no.~12, p. e0208561, 2018.

\bibitem{chu2023qdoor}
C.~Chu and et~al., ``Qdoor: Exploiting approximate synthesis for backdoor attacks in quantum neural networks,'' in \emph{2023 IEEE International Conference on Quantum Computing and Engineering (QCE)}, vol.~1.\hskip 1em plus 0.5em minus 0.4em\relax IEEE, 2023.

\bibitem{john2025quantum}
J.~John, L.~Golla, and Q.~Wang, ``Quantum trojan insertion: Controlled activation for covert circuit manipulation,'' \emph{arXiv preprint arXiv:2502.08880}, 2025.

\bibitem{liu2018trojaning}
Y.~Liu and et~al., ``Trojaning attack on neural networks,'' in \emph{25th Annual Network And Distributed System Security Symposium (NDSS 2018)}.\hskip 1em plus 0.5em minus 0.4em\relax Internet Society, 2018.

\bibitem{kundu2024adversarial}
S.~Kundu and S.~Ghosh, ``Adversarial poisoning attack on quantum machine learning models,'' \emph{arXiv preprint arXiv:2411.14412}, 2024.

\bibitem{chu2023qtrojan}
C.~Chu and et~al., ``Qtrojan: A circuit backdoor against quantum neural networks,'' in \emph{ICASSP 2023-2023 IEEE International Conference on Acoustics, Speech and Signal Processing (ICASSP)}.\hskip 1em plus 0.5em minus 0.4em\relax IEEE, 2023.

\bibitem{upadhyay2023stealthyswapsadversarialswap}
\BIBentryALTinterwordspacing
S.~Upadhyay and S.~Ghosh, ``Stealthy swaps: Adversarial swap injection in multi-tenant quantum computing,'' 2023. [Online]. Available: \url{https://arxiv.org/abs/2310.17426}
\BIBentrySTDinterwordspacing

\bibitem{reers2024comparative}
V.~Reers and M.~Mau{\ss}ner, ``Comparative analysis of vulnerabilities in classical and quantum machine learning,'' in \emph{INFORMATIK 2024}.\hskip 1em plus 0.5em minus 0.4em\relax Gesellschaft f{\"u}r Informatik eV, 2024, pp. 555--571.

\bibitem{nielsen2010quantum}
M.~A. Nielsen and I.~L. Chuang, \emph{Quantum computation and quantum information}.\hskip 1em plus 0.5em minus 0.4em\relax Cambridge university press, 2010.

\bibitem{schuld2020circuit}
M.~Schuld, A.~Bocharov, K.~M. Svore, and N.~Wiebe, ``Circuit-centric quantum classifiers,'' \emph{Physical Review A}, vol. 101, no.~3, p. 032308, 2020.

\bibitem{foulds2021controlled}
S.~Foulds, V.~Kendon, and T.~Spiller, ``The controlled swap test for determining quantum entanglement,'' \emph{Quantum Science and Technology}, vol.~6, no.~3, p. 035002, 2021.

\bibitem{liu2024quantum}
W.~Liu, Y.-Z. Li, H.-W. Yin, Z.-R. Wang, and J.~Wu, ``Quantum multi-state swap test: an algorithm for estimating overlaps of arbitrary number quantum states,'' \emph{EPJ Quantum Technology}, vol.~11, no.~1, p.~46, 2024.

\bibitem{maldonado2024quantum}
A.~Maldonado-Romo, J.~Y. Montiel-P{\'e}rez, V.~Onofre, J.~Maldonado-Romo, and J.~H. Sossa-Azuela, ``Quantum k-nearest neighbors: Utilizing qram and swap-test techniques for enhanced performance,'' \emph{Mathematics}, vol.~12, no.~12, p. 1872, 2024.

\bibitem{kay2018tutorial}
A.~Kay, ``Tutorial on the quantikz package,'' \emph{arXiv preprint arXiv:1809.03842}, 2018.

\bibitem{ripper2023swap}
P.~Ripper, G.~Amaral, and G.~Tempor{\~a}o, ``Swap test-based characterization of decoherence in universal quantum computers,'' \emph{Quantum Information Processing}, vol.~22, no.~5, p. 220, 2023.

\bibitem{el2024advqunn}
W.~El~Maouaki, A.~Marchisio, T.~Said, M.~Bennai, and M.~Shafique, ``Advqunn: A methodology for analyzing the adversarial robustness of quanvolutional neural networks,'' in \emph{2024 IEEE International Conference on Quantum Software (QSW)}.\hskip 1em plus 0.5em minus 0.4em\relax IEEE, 2024, pp. 175--181.

\bibitem{el2024robqunns}
W.~El~Maouaki, A.~Marchisio, T.~Said, M.~Shafique, and M.~Bennai, ``Robqunns: A methodology for robust quanvolutional neural networks against adversarial attacks,'' in \emph{2024 IEEE International Conference on Image Processing Challenges and Workshops (ICIPCW)}.\hskip 1em plus 0.5em minus 0.4em\relax IEEE, 2024, pp. 4090--4095.

\bibitem{lecun1998gradient}
Y.~LeCun, L.~Bottou, Y.~Bengio, and P.~Haffner, ``Gradient-based learning applied to document recognition,'' \emph{Proceedings of the IEEE}, vol.~86, no.~11, pp. 2278--2324, 1998.

\bibitem{krizhevsky2014cifar}
A.~Krizhevsky, V.~Nair, G.~Hinton \emph{et~al.}, ``The cifar-10 dataset,'' \emph{online: http://www. cs. toronto. edu/kriz/cifar. html}, vol.~55, no.~5, p.~2, 2014.

\bibitem{paszke2019pytorch}
A.~Paszke, ``Pytorch: An imperative style, high-performance deep learning library,'' \emph{arXiv preprint arXiv:1912.01703}, 2019.

\bibitem{neumaier2019born}
A.~Neumaier, ``Born's rule and measurement,'' \emph{arXiv preprint arXiv:1912.09906}, 2019.

\bibitem{ghosh2024ai}
A.~Ghosh and S.~Ghosh, ``Ai-driven reverse engineering of qml models,'' \emph{arXiv preprint arXiv:2408.16929}, 2024.

\bibitem{abreu2024qml}
D.~Abreu, C.~E. Rothenberg, and A.~Abel{\'e}m, ``Qml-ids: Quantum machine learning intrusion detection system,'' in \emph{2024 IEEE Symposium on Computers and Communications (ISCC)}.\hskip 1em plus 0.5em minus 0.4em\relax IEEE, 2024, pp. 1--6.

\bibitem{li2016survey}
H.~Li, Q.~Liu, and J.~Zhang, ``A survey of hardware trojan threat and defense,'' \emph{Integration}, vol.~55, pp. 426--437, 2016.

\end{thebibliography}

\end{document}